\documentclass[prl,reprint,10pt]{revtex4-1}
\usepackage{amssymb}
\usepackage{graphicx,float,amsmath}
\usepackage{amssymb,amsmath,graphicx}
\usepackage[usenames,dvipsnames]{color}
\usepackage{array}
\definecolor{MyDarkBlue}{rgb}{0.15,0.15,0.45}
\usepackage[linktocpage=true]{hyperref}
\hypersetup{
colorlinks=true,
citecolor=MyDarkBlue,
linkcolor=MyDarkBlue,
urlcolor=MyDarkBlue,
pdfauthor={},
pdftitle={},
pdfsubject={}
}

\allowdisplaybreaks

\newcommand{\be}{\begin{equation}}
\newcommand{\ee}{\end{equation}}
\newcommand{\sbe}{\begin{subequations}}
\newcommand{\see}{\end{subequations}}
\def\ba{\begin{align}}
\def\ea{\end{align}}
\newcommand{\p}{\partial}
\newcommand{\nn}{\nonumber}
\newcommand{\ud}{\mathrm{d}}
\newcommand{\calQ}{\mathcal{Q}}
\newcommand{\calE}{\mathcal{E}}

\begin{document}

\title{Dipolar tidal effects in scalar-tensor theories}
\author{Laura Bernard} 
\email{Laura.Bernard@obspm.fr}
\affiliation{LUTH, Observatoire de Paris, PSL Research University, CNRS, Universit\'e Paris Diderot, Sorbonne Paris Cit\'e, 5 place Jules Janssen, 92195 Meudon, France}
\date{\today}

\def\lgrho{\log_{10}\rho}

\begin{abstract}

The inclusion of finite-size effects in the  gravitational waveform templates allows one not only to constrain the internal structure of compact objects, but to test deviations from general relativity. Here, we address the problem of tidal effects in massless scalar-tensor theories. We introduce the scalar-type tidal Love numbers that relate the time-varying scalar dipole moment to the induced scalar tidal field. We compute the leading-order scalar tidal contribution in the conservative dynamics and for the first time in the wave generation for quasi-circular orbits. Importantly, we show that, in a system dominated by dipolar emission, such tidal effects may be detectable by third generation detectors such as the Einstein Telescope.

\end{abstract}

\maketitle

\paragraph{\bf\textit{Introduction.}}

The recent detections of gravitational waves by the LIGO-Virgo observatory have opened a new window to test the strong regime of gravity~\cite{Barack:2018yly}. In particular, the first multi-messenger observation of the coalescence of a neutron star (NS) binary system GW170817~\cite{TheLIGOScientific:2017qsa} has already constrained the NS equation of state~\cite{Abbott:2018exr}, which contains most of the information about the structure of the ultra dense nuclear matter inside neutron stars. Our ability to put more stringent constraints with the future space-based interferometer LISA and third-generation detectors relies on the modeling of the finite-size effects within the gravitational waveform templates. In particular, the precise incorporation of tidal effects requires solving the problem of the influence of NS nuclear physics on the gravitational signal. This is already a challenge in general relativity (GR) as the tidal effects start contributing at 5PN order~\footnote{As usual, we refer to post-Newtonian order as ${n\mathrm{PN}=\mathcal{O}\left(\frac{v^{2n}}{c^{2n}}\right)}$.} beyond the leading order, and add to other parameter effects in the model, due to the masses, spins, etc.

In a binary system, the external tidal field sourced by the companion induces a change in the orbital motion and a quadrupolar oscillation of the other object. In the adiabatic approximation in GR, this excitation is described by the ratio $\lambda$ between the induced quadrupole and the external tidal field. Finite size effects are expected to be very small during the inspiral phase of the coalescence, and the effect of the internal structure of the objects will only affect significantly the signal during the late inspiral and merger. However, one can still obtain information on the internal structure of neutron stars in the inspiral phase, due to our ability to model these effects in a very simple and clean manner that depends only on the dimensionless quantity defined from $\lambda$, called the tidal Love number (TLN). The general relativistic tidal contribution to the waveform was calculated for the first time in~\cite{Flanagan:2007ix,Hinderer:2007mb}, and extended to higher multipole moments~\cite{Binnington:2009bb,Damour:2009vw}. The tidal Love numbers are divided into the electric TLNs that correspond to the mass multipole moments and the magnetic TLNs linked to the current multipole moments. These effects are now properly modelled in GR up to the next-to-leading order (\textit{i.e.} 6PN)~\cite{Vines:2011ud,Yagi:2013sva,Banihashemi:2018xfb}, and they have been partially incorporating in the effective-one-body waveforms up to 7.5PN~\cite{Bini:2012gu,Damour:2012yf}.

The current work is part of the on-going effort on massless scalar-tensor (ST) theory to build the full gravitational and scalar waveforms at 2PN order beyond the leading GR order. Adding a single massless scalar field, minimally coupled to gravity, is one of the simplest ways to extend general relativity, motivated by cosmological observations. Currently, the tensor waveform is known at 2PN order~\cite{Lang:2013fna}, while the scalar waveform and the energy flux are, respectively known at 1.5PN and 1PN order~\cite{Lang:2014osa}.
To have a complete result at 2PN order for all quantities, one needs to use the dynamics at the higher 3PN order~\cite{Bernard:2018hta,Bernard:2018ivi}.  Finally, to complete these results one also has to include the finite-size effects. Indeed, due to the presence of a scalar dipole moment, the tidal effects start as early as the 3PN order, making it more easily accessible by current and forthcoming gravitational wave observations. Here, we aim to compute for the first time the scalar tidal effects that arise due to the presence of a scalar field coupled to gravity. In ST theories, in addition to the usual mass and current multipole moments, there are also some scalar multipole moments. We introduce a new class of TLNs, the scalar ones, to relate the time-varying scalar dipole moment to the induced scalar tidal field~\footnote{The scalar TLNs were first mentionned in Ref.~\cite{Cardoso:2017cfl}.}. In the following, we compute the leading order scalar tidal contribution to the dynamics and for the first time to the phase in the waveform. Our main result is that, when the binary system is dominated by dipolar emission, the tidal corrections for frequencies $\sim 0.01-1$Hz may contribute by $\mathcal{O}(1)$ to the waveform and by $\mathcal{O}(100)$ cycles for frequencies  $\sim 10-10^{3}$Hz, making their detection reachable by the future detectors LISA or third generation. While we only focus on the tidal effects originating in the scalar field, we leave the inclusion of the quadrupolar tidal effects to future work as they only start contributing at 5PN order.

\paragraph{\bf\textit{Scalar tidal Love number.}}

To motivate our treatment of the tidal interaction, we start by looking at the Newtonian theory. In the presence of a scalar field, $\phi$, the tidal field of the companion object will excite the dipole moment of a star. For a neutron star, the dipolar oscillation can be described by a tidally-driven harmonic oscillator, with no damping coefficient (as we neglect the viscosity)~\cite{Lai:1993di}. The corresponding Lagrangian is
\be\label{Lt}
L_{\rm{T}}=\frac{1}{2\lambda_{(s)}\omega_{1}^{2}}\left[\dot{\calQ}_{(s)}^{i}\dot{\calQ}^{(s)}_{i}-\omega_{1}^{2}\calQ_{(s)}^{i}\calQ_{i}^{(s)}\right]-\calE_{i}^{(s)}\calQ_{(s)}^{i} \,,
\ee
where $\calQ_{(s)}^{i}$ is the internal dynamical dipole, $\calE_{i}^{(s)}$ is the external dipolar tidal field sourced by the scalar field, $\omega_{1}$ is the frequency of the $l=1$ fundamental oscillating mode of the neutron star and $\lambda_{(s)}$ is the scalar tidal deformability parameter. Varying the action with respect to $\calQ^{i}_{(s)}$, we obtain the equation for a tidally-driven harmonic oscillator $\ddot{\calQ}_{(s)}^{i}+\omega_{1}^{2}\calQ_{(s)}^{i}=-\lambda_{(s)}\omega_{1}^{2}\calE_{(s)}^{i}$. As we are interested in the limit when the oscillator evolves adiabatically, we consider the stationary solution $\calQ_{(s)}^{i} = -\lambda_{(s)} \calE_{(s)}^{i}$.
By reinjecting it into the Lagrangian~\eqref{Lt}, we get the Newtonian correction to the Lagrangian in the presence of an external tidal field, $L_{\rm{T}}=\frac{1}{2\lambda_{(s)}}\calE_{i}^{(s)}\calE_{(s)}^{i}$.

By analogy with the Newtonian case, we now write a general relativistic theory of tidal interaction in the presence of a scalar field. The dipole moment induced by the external scalar tidal field $\calE^{(s)}_{\mu}\equiv\p_{\mu}\phi$, is
\be
\calQ^{(s)}_{\mu} = -\lambda_{(s)} \calE^{(s)}_{\mu}\,.
\ee
The coefficient of proportionality $\lambda_{(s)}$ is the scalar tidal deformability parameter and has a dimension of ${[\frac{G\lambda_{(s)}}{c^{2}}]=[\mathrm{length}]^{3}}$. Alternatively, one can define a dimensionless quantity, called the scalar tidal Love number (sTLN), ${k^{(s)}\equiv\frac{c^{4}}{G^{2}M_{A}^{3}}\lambda^{(s)}\sim\frac{G}{c^{2}R_{A}^{3}}\lambda^{(s)}}$, where we have used $\frac{G M_{A}}{c^{2}R_{A}}\sim 1$, with $M_{A}$ the mass of body A and $R_{A}$ its radius~\footnote{For a discussion on the gauge independence of such Love numbers, see~\cite{Binnington:2009bb}.}. Finally, as already proposed in Ref.~\cite{Damour:1998jk}, the action describing the finite-size effects is
\be\label{DSfsInit}
\Delta S_{(fs)} = -\sum_{A}\frac{1}{2}\lambda_{A}^{(s)}\int\!\!\ud s_{A}c\,(g^{\mu\nu})_{A}(\p_{\mu}\phi)_{A}(\p_{\nu}\phi)_{A}\,.
\ee
Note that this interaction is absent in general relativity as the dipole moment created by the Newtonian potential is zero in the center-of-mass frame.

\paragraph{\bf\textit{Contribution to the dynamics.}}

We consider the scalar-tensor action describing a massless scalar field $\phi$ coupled to the metric $g_{\mu\nu}$~\footnote{The action~\eqref{SstTide} does not contain self-interacting terms as we focus on massless ST theories. For a discussion on the phenomenology of massive ST theories, see Refs.~\cite{Alsing:2011er,Huang:2018pbu,Capozziello:2008rq}.},
\be\label{SstTide}
S \!= \!\frac{c^{3}}{16\pi G} \!\int\!\!\ud^{4}x\sqrt{-g}\!\left[\phi R - \frac{\omega(\phi)}{\phi}\nabla^{\alpha}\phi\nabla_{\alpha}\phi\right] +S_{\rm{pp}} +\Delta S_{(fs)} ,
\ee
where $S_{\rm{pp}}=-\sum_{A}c\int\!\ud s_{A} m_{A}(\phi)$ is the skeletonized action describing the bodies as point particles. The internal self-gravity of the compact objects is incorporated through a scalar-field dependent mass $m_{A}(\phi)$~\cite{Eardley1975}. We then define the sensibility of each object ${s_{A}\equiv\left.\frac{\ud\ln m_{a}(\phi)}{\ud\ln\phi}\right|_{0}}$, the subscript meaning that the scalar field is evaluated at spatial infinity, $\phi_{0}=\phi(\infty)$, where it is assumed to be constant in time. It measures how the internal structure of the body is affected by the presence of the scalar field. The last term in Eq.~\eqref{SstTide} incorporates the finite-size effects and, according to the previous section, is given by
\be\label{DSfs}
\Delta S_{(fs)} = -\sum_{A}\frac{1}{2}\lambda_{A}^{(s)}\int\!\!\ud s_{A}c\,(g^{\mu\nu})_{A}(\p_{\mu}\varphi)_{A}(\p_{\nu}\varphi)_{A}\,,
\ee
where we have defined $\varphi\equiv\frac{\phi}{\phi_{0}}$. Similarly to the mass of point-particles in ST theories, the scalar tidal Love numbers should be viewed as a function of the scalar field. However, as we are interested only in the main contribution, we consider $\lambda_{A}^{(s)}$ as being the zeroth order in an expansion in the scalar field, and thus constant. In order to compute the leading order tidal contribution to the dynamics of the system, we implement the post-Newtonian formalism~\cite{Blanchet:2013haa}. After deriving the Euler-Lagrange equations from the total action~\eqref{SstTide}, we solve the scalar and gravitational equations of motion for the metric iteratively at each PN order~\cite{Bernard:2018hta}. The equations of motion for the particles are then obtained from the geodesic equations. The Newtonian dynamics is the same as in GR, but with an effective coupling constant $\tilde{G}\alpha$, where $\tilde{G}=\frac{G(4+2\omega_{0})}{\phi_{0}(3+2\omega_{0})}$ and $\alpha=1-\zeta+\zeta(1-2s_{1})(1-2s_{2})$, with $\zeta=\frac{1}{4+2\omega_{0}}$ and $\omega_{0}$ is the value of  $\omega(\phi)$ at infinity. At higher orders, new parameters have to be introduced, \textit{i.e.} $\overline{\delta}_{A}\equiv\frac{\zeta(1-\zeta)}{\alpha^{2}}(1-2s_{A})^2$ and $\overline{\gamma}\equiv-\frac{2\zeta}{\alpha}(1-2s_{1})(1-2s_{2})$~\cite{Bernard:2018hta}.

As we are interested in the finite size contributions, we only have to compute the corrections to the Newtonian dynamics due to the dipolar tidal coupling.  In particular, the scalar field equation is modified with two new source terms,
\begin{align}\nn
\square_{g}\phi = &\frac{1}{3+2\omega}\Biggl[\frac{8\pi G}{c^{4}\phi_{0}}\left(T+\Delta T\right) \\
&\ - \frac{16\pi G}{c^{4}}\phi\left(\frac{\partial T}{\partial\phi}+\Delta S\right)-\omega' \nabla^{\lambda}\phi\nabla_{\lambda}\phi\Biggr]\,,
\end{align}
where a prime indicates derivation with respect to $\phi$, and $\Delta T^{\mu\nu} = \frac{2}{\sqrt{-g}}\frac{\delta\Delta S_{(\mathrm{fs})}}{\delta g_{\mu\nu}}$, $T\equiv g_{\mu\nu}T^{\mu\nu}=\frac{2g_{\mu\nu}}{\sqrt{-g}}\frac{\delta S_{(\mathrm{pp})}}{\delta g_{\mu\nu}}$, $\Delta T\equiv g_{\mu\nu}\Delta T^{\mu\nu}$ and $\Delta S = \frac{1}{\sqrt{-g}}\frac{\delta\Delta S_{(\mathrm{fs})}}{\delta \phi}$. Similarly the geodesic equations can now be written in the form ${\nabla_{\nu}\left(T^{\mu\nu}+\Delta T^{\mu\nu}\right) = \left(\frac{\partial T}{\partial\phi}+\Delta S\right)\nabla^{\mu}\phi}$. Note that we should be careful when treating this equation as all the terms have to be taken in the sense of distributions. Their contribution to the relative acceleration $\mathbf{a}\equiv\mathbf{a}_{1}-\mathbf{a}_{2}$ is given by
\begin{align}\label{daifs}
\Delta \mathbf{a}^{(fs)} = -\frac{\tilde{G}\alpha m}{r^{3}}\mathbf{x} \cdot\frac{-8 \zeta}{1-\zeta}\biggl[ \frac{m_{2}}{m_{1}}\,\overline{\delta}_{2}\,\lambda_{1}^{(s)} + \frac{m_{1}}{m_{2}}\,\overline{\delta}_{1}\,\lambda_{2}^{(s)}\biggr]\frac{\tilde{G}\alpha}{c^2\,r^3} .
\end{align}
where $\mathbf{x}=\mathbf{y_{1}}-\mathbf{y_{2}}$ is the relative position and $m\equiv m_{1}+m_{2}$ is the total mass of the system.
From this expression and using the sTLN $k_{A}^{(s)}\equiv\lambda_{A}^{(s)}\frac{\tilde{G}\alpha}{c^{2}R_{A}^{3}}$, we see that the tidal correction to the equations of motion are of 3PN order.
Then, as we also have $\frac{\tilde{G}\alpha M_{A}}{c^{2}R_{A}}\sim 1$ for compact objects in ST theories, the relative acceleration scales as the third power of the radius $R_{A}$. It implies that the impact on the dynamics may be enhanced for a less compact star compared to the naive 3PN expectation. Note that for a neutron-star black hole system, the tidal correction is zero, as black holes have zero scalar charge and are expected to have a vanishing scalar Love number~\cite{Cardoso:2017cfl}.

Performing a Legendre transform from the Lagrangian, we get the tidal contributions to the conserved energy for circular orbits,
\begin{align}\nn
\Delta E_{(fs)}(x) = -\frac{1}{2}m\nu c^{2}x \cdot & \frac{20\zeta}{3(1-\zeta)}\biggl[\frac{m_{2}}{m_{1}}\,\overline{\delta}_{2}\,\lambda_{1}^{(s)} \\
&\qquad + \frac{m_{1}}{m_{2}}\,\overline{\delta}_{1}\,\lambda_{2}^{(s)}\biggr]\frac{\tilde{G}\alpha}{c^2\,r^3}\,,
\end{align}
where we have introduced the symmetric mass ratio $\nu\equiv\frac{m_{1}m_{2}}{m^2}$, the PN parameter $x\equiv\left(\frac{\tilde{G}\alpha m\omega}{c^{3}}\right)^{2/3}$, $\omega$ beigin the orbital frequency, and $r$ on the right-hand side has to be understood as a function of x, \textit{i.e.} $r=r_{\mathrm{N}}=\frac{\tilde{G}\alpha m}{c^2 x}$. This is consistent with the early result on finite-size effect in scalar-tensor theories obtained in~\cite{Damour:1998jk}.

\paragraph{\bf\textit{Effect on the gravitational wave signal.}}

We now use the previous results to express the waveform in terms of the orbital phase $\psi$ and frequency $\omega$. The time derivative of the energy is computed from the energy flux through the balance equation, $\frac{\ud E_{S}}{\ud t}=-\mathcal{F}$, and is given by the formula~\cite{Lang:2014osa}
\be\label{EnergyFlux}
\frac{\ud E_{S}}{\ud t} = -\frac{G\phi_{0}(3+2\omega_{0})}{3c^{3}}\,\ddot{I}^{(s)}_{i}\,\ddot{I}_{(s)}^{i}\,,
\ee
where dotted quantities are derived with respect to time. The scalar dipole moment has to be corrected by tidal effects, $I_{(s)}^{i}(t) = I_{(s), \mathrm{N}}^{i}+I_{(s), \mathrm{fs}}^{i}$, with~\cite{Bernard:2018hta}
\sbe\begin{align}
I_{(s), \mathrm{N}}^{i} & = \frac{2m\nu (s_{1}-s_{2})}{\phi_{0}(3+2\omega_{0})}\,x^{i} \,,\\
\nn I_{(s), \mathrm{fs}}^{i} & = -\frac{4G}{c^{2}\phi_{0}^2(3+2\omega_{0})^2}\Bigl[m_{2}(1-2s_{2})\lambda_{1}^{(s)} \\
&\qquad\qquad\qquad\qquad - m_{1}(1-2s_{1})\lambda_{2}^{(s)}\Bigr]\,\frac{x^{i}}{r^{3}}\,.
\end{align}\see
Incorporating the dipole moment into the formula~\eqref{EnergyFlux}, we get
\begin{widetext}\begin{align}\nn
\left.\frac{\ud E_{S}}{\ud t}\right|_{(fs)} =\ & -\frac{4\nu^2}{3\tilde{G}\alpha} \frac{\zeta(s_{1}-s_{2})^{2}}{\alpha} c^{5}x^{4} \cdot\Biggl[ -\frac{16\zeta}{3(1-\zeta)}\left(\frac{m_{1}}{m_{2}}\overline{\delta}_{1}\left(\frac{R_{2}}{r}\right)^{3}k_{2}^{(s)}+\frac{m_{2}}{m_{1}}\overline{\delta}_{2}\left(\frac{R_{1}}{r}\right)^{3}k_{1}^{(s)}\right) \\
&\qquad\ -8\frac{\zeta}{\alpha}\left(\left(1+\frac{m_{1}}{m_{2}}\right)\frac{\overline{\delta}_{1}+\frac{\overline{\gamma}(2+\overline{\gamma})}{4}}{\overline{\delta}_{1}+\overline{\delta}_{2}+\frac{\overline{\gamma}(2+\overline{\gamma})}{2}}\left(\frac{R_{2}}{r}\right)^{3}k_{2}^{(s)}+\left(1+\frac{m_{2}}{m_{1}}\right)\frac{\overline{\delta}_{2}+\frac{\overline{\gamma}(2+\overline{\gamma})}{4}}{\overline{\delta}_{1}+\overline{\delta}_{2}+\frac{\overline{\gamma}(2+\overline{\gamma})}{2}}\left(\frac{R_{1}}{r}\right)^{3}k_{1}^{(s)}\right) \Biggr]\,,
\end{align}\end{widetext}
where again $r$ has to be understood as $r_{\mathrm{N}}=\frac{\tilde{G}\alpha m}{c^2 x}$ and we have factorized the leading contribution.
Then, using the definition of the PN parameter $x=\left(\frac{\tilde{G}\alpha m\omega}{c^{3}}\right)^{2/3}$ and the relation $\dot{\psi}=\omega$, we get the formula ${\frac{\ud \psi}{\ud x} = \frac{(c^{2}x)^{3/2}}{\tilde{G}\alpha m}\frac{\ud E/\ud x}{\ud E/\ud t}}$. In order to obtain the leading order tidal correction to the phase, we re-expand $\frac{\ud \psi}{\ud x}$ in $x$ and integrate term by term. This procedure corresponds to the TaylorT2 approximant method that gives an analytic result~\cite{Buonanno:2009zt}.

Before going further, we should look carefully at the expansions we are performing. Indeed, the scalar-tensor parameters are already strongly constrained by solar-system and pulsar observations, resulting in a correction smaller than expected~\cite{Freire:2012mg}. To study the balance between the orbital parameters and the ST ones, we compute the ratio between the dipolar and quadrupolar energy fluxes~\cite{Sennett:2016klh},
\be
\frac{\mathcal{F}_{\mathrm{dip}}}{\mathcal{F}_{\mathrm{quad}}} = \frac{5\zeta(s_{1}-s_{2})^{2}}{24\bigl(1-\zeta+\frac{\zeta}{6}(1-s_{1}-s_{2})\bigr)x} \,.
\ee
We see that at low frequency or for systems with asymmetric large scalar charges, for example that undergo dynamical scalarization, the dipolar emission could dominate. However as was outlined in~\cite{Sennett:2016klh}, most of the systems that could be seen by the current detectors, LIGO and Virgo, or by the space-based one LISA will be dominated by the quadrupolar emission. Following~\cite{Sennett:2016klh}, we now separate the dipole-driven (DD) regime from the quadrupole-driven (QD) one. In the former case the energy flux is dominated by the dipolar term ${\cal F}_{\mathrm{dip}}$, while in the latter case the Newtonian term ${\cal F}_{\mathrm{quad}}$ dominates due to the smallness of the scalar-tensor parameters compared to the GW frequency.

\textit{i) Dipolar-driven regime.}
In the dipolar-driven regime, the energy flux is expanded around the dipolar flux, $\mathcal{F}_{DD}=\mathcal{F}_{\mathrm{dip}}+\left.\Delta\mathcal{F}\right|_{(\mathrm{fs})}$. The calculations are straightforward and we obtain the differential equation for the phase $\psi$
\begin{widetext}\begin{align}\label{dpsidxfsdd}\nn
\left.\frac{\ud\psi}{\ud x}\right|_{DD} =\ & \frac{3\alpha}{8\nu\zeta(s_{1}-s_{2})^{2}}x^{-5/2} \cdot\Biggl[1 +\frac{32\zeta}{(1-\zeta)}\left(\frac{m_{1}}{m_{2}}\overline{\delta}_{1}\left(\frac{R_{2}}{r}\right)^{3}k_{2}^{(s)}+\frac{m_{2}}{m_{1}}\overline{\delta}_{2}\left(\frac{R_{1}}{r}\right)^{3}k_{1}^{(s)}\right) \\
&\qquad\ +\frac{8\zeta}{\alpha}\left(\left(1+\frac{m_{1}}{m_{2}}\right)\frac{\overline{\delta}_{1}+\frac{\overline{\gamma}(2+\overline{\gamma})}{4}}{\overline{\delta}_{1}+\overline{\delta}_{2}+\frac{\overline{\gamma}(2+\overline{\gamma})}{2}}\left(\frac{R_{2}}{r}\right)^{3}k_{2}^{(s)}+\left(1+\frac{m_{2}}{m_{1}}\right)\frac{\overline{\delta}_{2}+\frac{\overline{\gamma}(2+\overline{\gamma})}{4}}{\overline{\delta}_{1}+\overline{\delta}_{2}+\frac{\overline{\gamma}(2+\overline{\gamma})}{2}}\left(\frac{R_{1}}{r}\right)^{3}k_{1}^{(s)}\right) \Biggr]\,,
\end{align}\end{widetext}
that we then integrate term by term to get the phase as a function of the frequency $x$. In order to put our result in a form that could be used to constrain the theory with the current and future GW detectors, we present the phase in the Fourier domain. Using the stationary phase approximation~\cite{Tichy:1999pv}, the waveform can be written as $\mathcal{A}\mathrm{e}^{i\Psi}$. The phase in Fourier domain is $\Psi_{l,m}(f) = m\left(\psi(v)-\dfrac{1}{Gm\alpha}v^{3}t(v)\right)$, where $v=x^{1/2}$ is to be evaluated at the GW frequency $f$ and $v_f=(\pi \tilde{G}\alpha m f)^{1/3}$. The tidal correction to the Fourier domain phase is then
\begin{widetext}\begin{align}\label{psifsdd}\nn
& \left. \Delta\Psi_{l,m}(f)\right|_{DD} =\ \frac{\alpha}{8\nu\zeta(s_{1}-s_{2})^{2}}\left(\frac{c}{v}\right)^{3} \cdot\Biggl[\frac{-64\zeta}{(1-\zeta)}\left(\frac{m_{1}}{m_{2}}\overline{\delta}_{1}\left(\frac{R_{2}}{r}\right)^{3}k_{2}^{(s)}+\frac{m_{2}}{m_{1}}\overline{\delta}_{2}\left(\frac{R_{1}}{r}\right)^{3}k_{1}^{(s)}\right) \\
&\qquad\quad -\frac{16\zeta}{\alpha}\left(\left(1+\frac{m_{1}}{m_{2}}\right)\frac{\overline{\delta}_{1}+\frac{\overline{\gamma}(2+\overline{\gamma})}{4}}{\overline{\delta}_{1}+\overline{\delta}_{2}+\frac{\overline{\gamma}(2+\overline{\gamma})}{2}}\left(\frac{R_{2}}{r}\right)^{3}k_{2}^{(s)}+\left(1+\frac{m_{2}}{m_{1}}\right)\frac{\overline{\delta}_{2}+\frac{\overline{\gamma}(2+\overline{\gamma})}{4}}{\overline{\delta}_{1}+\overline{\delta}_{2}+\frac{\overline{\gamma}(2+\overline{\gamma})}{2}}\left(\frac{R_{1}}{r}\right)^{3}k_{1}^{(s)}\right) \Biggr]\left(1-3\ln\left(\frac{v}{c}\right)\right)\,.
\end{align}\end{widetext}
where $v=v_f=(\pi \tilde{G}\alpha m f)^{1/3}$.

\textit{ii) Quadrupolar-driven regime.} 
In the quadrupolar-driven regime, taking the prescription introduced in~\cite{Sennett:2016klh}, we split the flux in the non-dipolar and the dipolar parts by defining,
\sbe\begin{align}
& \mathcal{F}_{\mathrm{nd}} \equiv \lim_{s_{1}-s_{2}\rightarrow 0} \mathcal{F}_{QD}  \,,\\
& \mathcal{F}_{\mathrm{dd}} \equiv \mathcal{F}_{QD}-\mathcal{F}_{\mathrm{nd}} \,,
\end{align}\see
and expand the flux around the leading non-dipolar term. Using the same decomposition for the other quantities, the differential equations for the phase are
\begin{widetext}\sbe\label{dpsidxfsnd}\begin{align}
\left.\frac{\ud\psi}{\ud x}\right|_{\mathrm{nd}} =\ & \frac{5\alpha}{64\nu\bigl(1-\zeta+\frac{\zeta}{6}(1-s_{1}-s_{2})\bigr)}x^{-7/2} \cdot\Biggl[1 +\frac{80\zeta}{3(1-\zeta)}\left(\frac{m_{1}}{m_{2}}\overline{\delta}_{1}\left(\frac{R_{2}}{r}\right)^{3}k_{2}^{(s)}+\frac{m_{2}}{m_{1}}\overline{\delta}_{2}\left(\frac{R_{1}}{r}\right)^{3}k_{1}^{(s)}\right)\Biggr] \,, \\
\nn \left.\frac{\ud\psi}{\ud x}\right|_{\mathrm{dd}} =\ & -\frac{25\alpha\zeta(s_{1}-s_{2})^{2}}{1536\nu\bigl(1-\zeta+\frac{\zeta}{6}(1-s_{1}-s_{2})\bigr)^2}x^{-9/2} \cdot\Biggl[1 -\frac{16\zeta}{3(1-\zeta)}\left(\frac{m_{1}}{m_{2}}\overline{\delta}_{1}\left(\frac{R_{2}}{r}\right)^{3}k_{2}^{(s)}+\frac{m_{2}}{m_{1}}\overline{\delta}_{2}\left(\frac{R_{1}}{r}\right)^{3}k_{1}^{(s)}\right) \\
&\qquad\ -\frac{8\zeta}{\alpha}\left(\left(1+\frac{m_{1}}{m_{2}}\right)\frac{\overline{\delta}_{1}+\frac{\overline{\gamma}(2+\overline{\gamma})}{4}}{\overline{\delta}_{1}+\overline{\delta}_{2}+\frac{\overline{\gamma}(2+\overline{\gamma})}{2}}\left(\frac{R_{2}}{r}\right)^{3}k_{2}^{(s)}+\left(1+\frac{m_{2}}{m_{1}}\right)\frac{\overline{\delta}_{2}+\frac{\overline{\gamma}(2+\overline{\gamma})}{4}}{\overline{\delta}_{1}+\overline{\delta}_{2}+\frac{\overline{\gamma}(2+\overline{\gamma})}{2}}\left(\frac{R_{1}}{r}\right)^{3}k_{1}^{(s)}\right) \Biggr]\,.
\end{align}\see\end{widetext}
As for the dipolar-driven regime, we compute the phase in the Fourier domain using the SPA and get the phase in the quadrupolar-driven regime,
\begin{widetext}\sbe\label{psifsnd}\begin{align}
\left.\Delta\Psi_{l,m}(f)\right|_{\mathrm{nd}} =\ & -\frac{3\alpha}{256\nu\bigl(1-\zeta+\frac{\zeta}{6}(1-s_{1}-s_{2})\bigr)}\left(\frac{c}{v}\right)^{5} \cdot\Biggl[\frac{-1600\zeta}{3(1-\zeta)}\left(\frac{m_{1}}{m_{2}}\overline{\delta}_{1}\left(\frac{R_{2}}{r}\right)^{3}k_{2}^{(s)}+\frac{m_{2}}{m_{1}}\overline{\delta}_{2}\left(\frac{R_{1}}{r}\right)^{3}k_{1}^{(s)}\right)\Biggr] \,, \\
\nn \left.\Delta\Psi_{l,m}(f)\right|_{\mathrm{dd}} =\ & \frac{5\alpha\zeta(s_{1}-s_{2})^{2}}{3584\nu\bigl(1-\zeta+\frac{\zeta}{6}(1-s_{1}-s_{2})\bigr)^2}\left(\frac{c}{v}\right)^{7} \cdot\Biggl[-\frac{280\zeta}{3(1-\zeta)}\left(\frac{m_{1}}{m_{2}}\overline{\delta}_{1}\left(\frac{R_{2}}{r}\right)^{3}k_{2}^{(s)}+\frac{m_{2}}{m_{1}}\overline{\delta}_{2}\left(\frac{R_{1}}{r}\right)^{3}k_{1}^{(s)}\right) \\
&\qquad\ -\frac{140\zeta}{\alpha}\left(\left(1+\frac{m_{1}}{m_{2}}\right)\frac{\overline{\delta}_{1}+\frac{\overline{\gamma}(2+\overline{\gamma})}{4}}{\overline{\delta}_{1}+\overline{\delta}_{2}+\frac{\overline{\gamma}(2+\overline{\gamma})}{2}}\left(\frac{R_{2}}{r}\right)^{3}k_{2}^{(s)}+\left(1+\frac{m_{2}}{m_{1}}\right)\frac{\overline{\delta}_{2}+\frac{\overline{\gamma}(2+\overline{\gamma})}{4}}{\overline{\delta}_{1}+\overline{\delta}_{2}+\frac{\overline{\gamma}(2+\overline{\gamma})}{2}}\left(\frac{R_{1}}{r}\right)^{3}k_{1}^{(s)}\right) \Biggr]\,,
\end{align}\see\end{widetext}
where $v=v_f=(\pi \tilde{G}\alpha m f)^{1/3}$.

\paragraph{\bf\textit{Discussion.}}

Considering the current bound on the ST parameters~\cite{Sennett:2016klh}, we can determine the order of magnitude estimate of the tidal corrections to the phase, Eqs.~\eqref{psifsdd} and \eqref{psifsnd}. As it also requires the computation of the scalar TLNs for specific NS equations of state, which is beyond the goal of this paper, we take them as being of order $\sim 0.1$, similar to the highest relativistic TLNs computed for GR~\footnote{Note that the l=2 scalar Love numbers were calculated in Ref~\cite{Pani:2014jra} for some scalar-tensor theories.}. In the quadrupolar driven regime, the scalar tidal effect is negligible and well below LISA or third generation detectors detectability range. On the contrary, when the binary system is dominated by dipolar emission, the tidal corrections for frequencies in the LISA band, $\sim 0.01-1$Hz, may contribute by $\mathcal{O}(1)$ to the waveform, comparable to the 1PN relative scalar-tensor correction. In the Earth-based detectors frequency band, $\sim 10-10^{3}$Hz, the tidal contribution may be even higher, of order $\mathcal{O}(100)$, for a binary neutron star system.
This shows that while the correction to the phase due to tidal effect is formally at 3PN order, the effect may be larger than the naive 3PN expectation due to the scaling in $\left(R_{A}/m\right)^{3}=\left(R_{A}/M_{A}\right)^{3}\times\left(M_{A}/m\right)^{3}$, which can be of order $10^2$ for neutron stars~\cite{Yagi:2013baa}. It means that a larger star will have a stronger scalar tidal effect on the waveform, that may be as strong as the point-particle 1PN effect.
Moreover as it does not behave as the leading-order GR or the 1PN ST contribution, it may be even easier to detect. This makes the detection of scalar tidal effects reachable by the forthcoming third generation detectors and LISA.
All these features make very promising the use of tidal effects to put further constraints on scalar-tensor theories. We emphasize that it is crucial to incorporate them in the future scalar-tensor waveform templates that will be used in LISA and third generation GW detectors as the Einstein Telescope. They should also be taken into consideration when devising tests of gravity and of the cosmological models with the GW observations~\cite{Capozziello:2017vdi}.

\paragraph{\bf\textit{Acknowledgments.}}

The author would like to thank Tanja Hinderer, Eric Poisson and Huan Yang for valuable discussions on this work, and Eric Poisson for a careful reading of the manuscript. She would also like to thank Eve Dones and Banafsheh Shiralilou for pointing out some typos in the first published version of the article.
This research was supported in part by Perimeter Institute for Theoretical Physics. Research at Perimeter Institute is supported by the Government of Canada through the Department of Innovation, Science and Economic Development Canada and by the Province of Ontario through the Ministry of Research, Innovation and Science.


\bibliographystyle{apsrev4}
\bibliography{references}


\end{document}